# Energy Dissipation in Monolayer MoS₂ Electronics

Eilam Yalon[1], Connor J. McClellan[1], Kirby K. H. Smithe[1], Miguel Muñoz Rojo[1], Runjie (Lily) Xu[1], Saurabh V. Suryavanshi[1], Alex J. Gabourie[1], Christopher M. Neumann[1], Feng Xiong[1,2], Amir B. Farimani[3], and Eric Pop[1,4,5,*]

[1]*Department of Electrical Engineering, Stanford University, Stanford, CA 94305, USA.* [2]*Present address: Department of Electrical & Computer Engineering, University of Pittsburgh, Pittsburgh, PA 15261, USA.* [3]*Department of Chemistry, Stanford University, Stanford, CA 94305, USA* [4]*Department of Materials Science & Engineering, Stanford University, Stanford, CA 94305, USA.* [5]*Precourt Institute for Energy, Stanford University, Stanford, CA 94305, USA.* [*]*E-mail: epop@stanford.edu*

**The advancement of nanoscale electronics has been limited by energy dissipation challenges for over a decade. Such limitations could be particularly severe for two-dimensional (2D) semiconductors integrated with flexible substrates or multi-layered processors, both being critical thermal bottlenecks. To shed light into fundamental aspects of this problem, here we report the first direct measurement of spatially resolved temperature in functioning 2D monolayer MoS₂ transistors. Using Raman thermometry we simultaneously obtain temperature maps of the device channel and its substrate. This differential measurement reveals the thermal boundary conductance (TBC) of the MoS₂ interface (14 ± 4 MWm⁻²K⁻¹) is an order magnitude larger than previously thought, yet near the low end of known solid-solid interfaces. Our study also reveals unexpected insight into non-uniformities of the MoS₂ transistors (small bilayer regions), which do not cause significant self-heating, suggesting that such semiconductors are less sensitive to inhomogeneity than expected. These results provide key insights into energy dissipation of 2D semiconductors and pave the way for the future design of energy-efficient 2D electronics.**

Keywords: *Energy dissipation, 2D semiconductors, thermal boundary conductance, Raman thermometry, MoS₂*

The performance of nanoelectronics is most often constrained by thermal challenges,[1, 2] memory bottlenecks,[3] and nanoscale contacts.[4] The former have become particular acute, with high integration densities leading to high power density, and numerous interfaces (e.g. between silicon, copper, $SiO_2$) leading to high thermal resistance. New applications and new form-factors call for dense vertical integration into multi-layer "high-rise" processors for high-performance computing,[3] or integration with poor thermal substrates like flexible plastics (of thermal conductivity 5x lower than $SiO_2$ and nearly 500x lower than silicon) for wearable computing.[5] These are the two most likely platforms for incorporating 2D semiconductors into electronics, yet very little is known about fundamental limits or practical implications of energy dissipation in these contexts.

At its most basic level, energy dissipation begins in the ultra-thin transistor channel and is immediately limited by the insulating regions and thermal resistance with the interfaces surrounding it. Herbert Kroemer's observation[6] that "the interface is the device" is remarkably apt for 2D semiconductors such as monolayer $MoS_2$. These have no bulk, and are thus strongly limited by their interfaces. For instance, even some of the best electrical contacts known today add >50% parasitic resistance to $MoS_2$ transistors when these are scaled to sub-100 nm dimensions.[7] Similarly, thermal interfaces may be expected to limit energy dissipation from 2D electronics, and their understanding is essential. Nevertheless, a key challenge is the need to differentiate heating of the sub-nanometer thin 2D material from its environment. Here, Raman spectroscopy holds a unique advantage,[8, 9] as the temperature of even a monolayer semiconductor can be distinguished from the material directly under (or above) it, if the Raman signatures are distinct.[10]

Figure 1a shows our typical device structure and measurement setup. We utilize high-quality $MoS_2$ films grown by chemical vapor deposition (CVD) on $SiO_2$, with Si substrates which serve as back-gates[11] (see Methods). Micron-scale channel dimensions are chosen to minimize power dissipation at the contacts (Supporting Information Section 1) and to obtain good spatial resolution. Some transistors are entirely monolayer (1L) and others contain small (<0.5 $\mu m^2$) bilayer (2L) regions,[11] as seen in Figures. 1b-d. In the main manuscript we focus on the latter, partly because they represent a more extreme case of material variability, and partly to reveal insight into energy dissipation at such 1L/2L interfaces. (Supporting Information Section 2 describes measurements of 1L exfoliated $MoS_2$, with similar results.)

Figure 1e displays the characteristic Raman peaks of a MoS$_2$ channel in thermal equilibrium and when power is applied ($P \approx 1$ mW/µm$^2$). The Raman peaks red-shift due to heating and phonon softening, which serves as the temperature marker (see Methods).[9, 12-14] Importantly, both the MoS$_2$ temperature and the Si substrate temperature (directly underneath the MoS$_2$ channel) are acquired simultaneously in this measurement since their Raman peaks are both measurable and spectrally resolved. This has not been previously implemented, to our knowledge, yet we find it is crucial to avoid the need for any assumptions regarding heat sinking from the Si substrate. The MoS$_2$ temperature is obtained from the out-of-plane A$_1$' mode to avoid uncertainty of strain effects on the in-plane E' mode, and additional corrections are described in Supporting Information Sections 3 and 4.

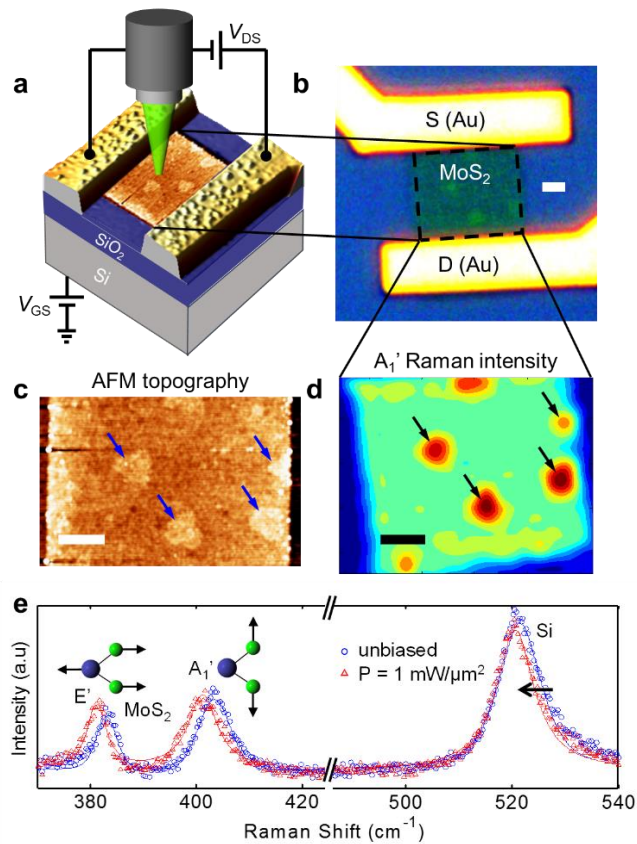

**Figure 1 | Raman thermometry of functioning monolayer MoS$_2$ transistor. (a)** Topography image overlaid on schematic device structure and experimental setup. The Raman signal is measured while electrical bias is applied. **(b)** Top view optical image of the device. **(c)** AFM topography image of the device. **(d)** Raman intensity map of the A$_1$' peak across the device. Small bilayer islands (marked by arrows) are visible in the optical image (b), AFM topography (c), and Raman map (d), but not in the temperature measurements (Figure 2). All scale bars are 1 µm. **(e)** Measured (symbols) and fitted (lines) Raman spectra at the center of the MoS$_2$ channel with electrical bias at $P = 1$ mW/µm$^2$ (red) and without bias (blue). Inset shows the atomic motions corresponding to the E' and A$_1$' Raman peaks of monolayer MoS$_2$.

The Raman peak shifts vs. temperature are first calibrated on a hot stage (Supporting Information Figure S5). Device temperature maps are then obtained by measuring the Raman peak shifts across the channel under electrical bias as shown in Figure 1e. Temperature maps of a MoS$_2$ transistor and their respective input power are shown in Figure 2a, revealing no temperature non-uniformities around the small 2L regions detected by AFM (Figures. 1c and 2c). The temperature uniformity of the device is confirmed by scanning thermal microscopy (SThM)[15] in Figure 2b and Supporting Information Figure S6. Unlike Raman, SThM only samples the temperature of the top AlO$_x$ capping layer (not the MoS$_2$ channel temperature), but the lack of temperature variation around 2L regions remains clearly evident. Similarly uniform temperature maps were obtained from exfoliated 1L devices, as shown in Supporting Information Figure S2. Minor, randomly distributed non-uniformities in the temperature seen in Figure 2 are within the uncertainty of the measurement and are also visible in the reference map taken at $V_{DS} = 0$ (on a hot stage), for which the temperature is known to be uniform, as shown in Supporting Information Figure S4.

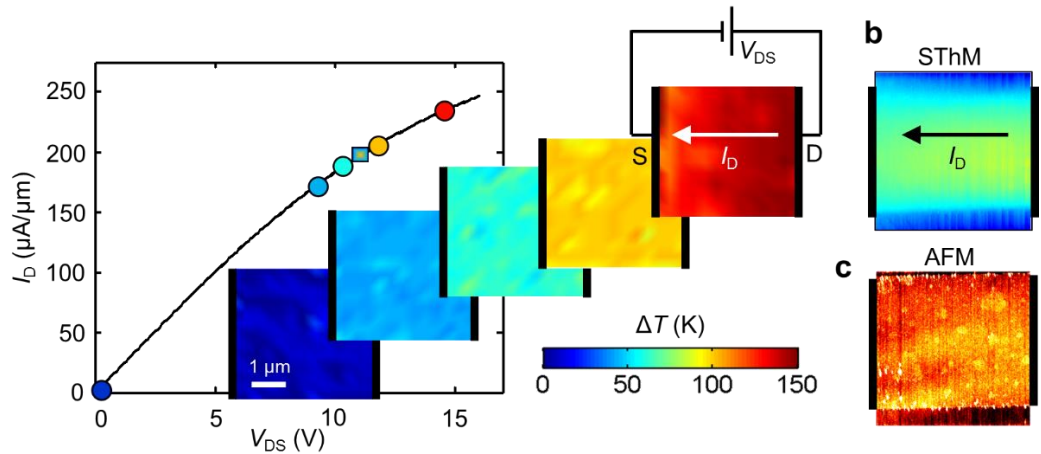

**Figure 2 | Measured temperature maps of MoS$_2$ transistor. (a)** Current vs. drain voltage and corresponding temperature maps, at back-gate $V_{GS} = 25$ V. Colored circles mark the bias point of each temperature color map. Arrows show current flow direction, right to left in all maps. At higher bias the heating becomes more significant at the drain side. However, temperature non-uniformities due to small bilayer regions (Figure 2c) are not observed. **(b)** SThM temperature map confirms relatively uniform temperature, despite small bilayer regions seen in **(c)** AFM topography. The AFM and SThM images are acquired simultaneously at the bias point marked with a square on the $I_D$–$V_{DS}$ curve. SThM evaluates the temperature at the top of the AlO$_x$ capping layer (see Methods), with values estimated from our thermal model, after calibration by Raman thermometry. Inset is temperature scale bar for all Raman and SThM color maps.

The uniform self-heating of transistors from CVD-grown $MoS_2$ suggests that any change in energy dissipation around the 2L spots or other non-uniformities is small, and below the resolution of the Raman thermometry technique. In fact, we utilize this information to place an upper bound on potential variations, like conduction band (CB) discontinuities at 1L-2L junctions, that could lead to measurable self-heating, and find these must be <120 meV (Supporting Information Figure S7). This finding is consistent with the previously estimated ~50 meV CB discontinuity at 1L-2L interfaces,[16, 17] and underscores that such 1L semiconductors are relatively immune to electrical variation introduced by small 2L regions which may occur during CVD growth. This CB variation is remarkably smaller than that expected of Si films with equivalent thickness variation between $d = 6.15$ Å and $2d$. The Si CB variation can be estimated from a simple quantum well model as $\Delta E_{CB} \sim 3h^2/(32m^*d^2) > 0.8$ eV, where $h$ is the Planck constant and $m^*$ is the effective mass in the Si CB,[18, 19] revealing that $MoS_2$ monolayers are much more immune to atomic-scale thickness variations than Si in this atomically thin limit.

Figure 3a shows the average temperature rise in the $MoS_2$ channel versus electrical input power density ($P$). No measurable difference is observed between CVD-grown (red) and exfoliated (blue) monolayer transistors, suggesting that their energy dissipation (and $MoS_2$-$SiO_2$ interface, as we will see below) is effectively the same. Importantly, our measurements simultaneously reveal the temperature rise at the underlying Si substrate surface (purple) directly beneath the $MoS_2$ channel. Knowledge of the Si temperature is essential to understand the energy dissipation and to validate the thermal model shown in Figure 3b,c.

The lines in Figure 3a represent the thermal resistance $\mathcal{R}_{th}$ normalized by the device area. The $\mathcal{R}_{th}$ ($= \Delta T_{MoS2}/P$) of the $MoS_2$ channel is the sum of contributions from the Si substrate ($\mathcal{R}_{th,Si}$ $= \Delta T_{Si}/P$), the $SiO_2$ layer ($\mathcal{R}_{th,ox}$), and the $MoS_2$-$SiO_2$ interface ($\mathcal{R}_{th,int}$), as illustrated in Figure 3c. This is a good approximation here, as the device dimensions are significantly larger than the lateral thermal healing length (~100 nm).[20, 21] We note that the $SiO_2$-Si interface TBC is > 125 $MWm^{-2}K^{-1}$, equivalent to < 10 nm Kapitza length in terms of $SiO_2$ thickness.[22, 23] This accounts for < 5% of $\mathcal{R}_{th}$, and is not shown in Figure 3c (see Supporting Information Section 8). The thermal resistance of the 90 nm thick $SiO_2$ is easily calculated because its thermal properties[22, 23] and the device dimensions are well known. Finally, since $\mathcal{R}_{th,Si}$ is directly measured, we can obtain the key thermal boundary conductance of the $MoS_2$-$SiO_2$ interface, TBC = $1/\mathcal{R}_{th,int} = 14 \pm 4$

MWm$^{-2}$K$^{-1}$. By comparison, molecular dynamics (MD) simulations of this interface yield TBC $\approx$ 15 MWm$^{-2}$K$^{-1}$ in good agreement with the experimental data (Supporting Information Section 9).

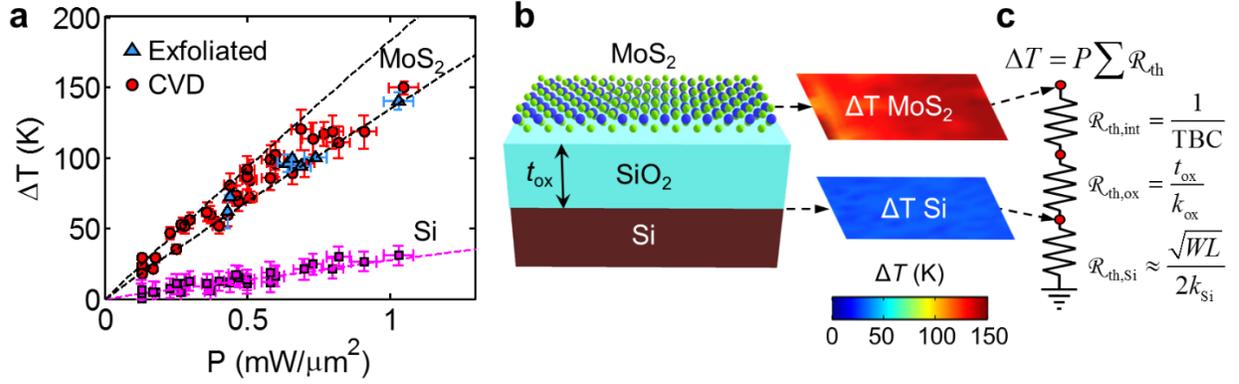

**Figure 3 | Device thermal resistance.** **(a)** Measured temperature rise of CVD-grown (red), exfoliated (blue) MoS$_2$ devices on SiO$_2$ ($t_{ox}$ = 90 nm) and their corresponding Si substrates (purple). Each marker represents averaged device temperature of 8 CVD and 3 exfoliated devices at varying input power, 42 total measurements. Black dashed lines denote the total thermal resistance $\mathcal{R}_{th}$ with TBC of 10 (upper) and 20 (lower) MWm$^{-2}$K$^{-1}$. Purple dashed line is the thermal resistance of the Si substrate, $\mathcal{R}_{th,Si}$. Error in power density is due to uncertainty of electrical contact resistance, while errors in measured temperature are due to Raman resolution and peak fitting uncertainty (see Supporting Information Sections 1 and 5). **(b)** Schematic cartoon of the device stack and simultaneously measured temperature rise maps of the MoS$_2$ channel and the Si surface directly underneath, as enabled by Raman thermometry. **(c)** Simple thermal model of 2D device, including the known temperature dependence of $k_{Si}(T)$ and $k_{ox}(T)$[21] when calculating the dashed lines in Figure 3a. The measured $\mathcal{R}_{th,Si}$ independently reveals the thermal conductivity of the substrate, $k_{Si} \approx (WL)^{1/2}/(2\mathcal{R}_{th,Si}) = 95 \pm 8$ Wm$^{-1}$K$^{-1}$, in good agreement with known values for highly doped Si.[24] The analytic term for thermal spreading resistance into the Si substrate is that of a circular disk heater, which is within < 5% error from the numerical solution of the rectangular transistor heat source (see Supporting Information Section 8).

The TBC found here is nearly an order of magnitude higher than recently reported for exfoliated 1L MoS$_2$ by Raman thermometry with optical heating,[12-14] but similar to that of metal interfaces with bulk MoS$_2$ (~25 MWm$^{-2}$K$^{-1}$).[25] The higher TBC cannot be explained solely by additional phonon coupling channels due to the presence of our AlO$_x$ capping layer,[26] but it could be due to better interface quality of our devices (see Methods). Our measurement accuracy is also improved by the precision of electrical heating power (used here for the first time to probe this interface), and our improved analysis which accounts for the thermal resistance of the SiO$_2$ while directly measuring the Si substrate (Figure 3a). In contrast, in optical heating experiments one must account for the temperature-dependent absorption, the precise laser spot size and shape, and for Raman shifts unrelated to temperature induced by high laser power. The latter are difficult to decouple from heating when the laser acts as both heater and thermometer.[12-14]

The agreement between our exfoliated and CVD-grown devices (as well as our MD simulations) also suggests that the TBC measured here likely approaches the upper limit of the "atomically intimate" interface. Nevertheless, we note that the $MoS_2$-$SiO_2$ TBC is near the very low end of known solid-solid interfaces (which range from ~10 $MWm^{-2}K^{-1}$ for Bi-diamond to 14 $GWm^{-2}K^{-1}$ for Pd-Ir),[2, 27] with a thermal resistance comparable to that of the underlying $SiO_2$ (~90 nm). This is an important result, because it highlights that energy dissipation from such 2D electronics is strongly limited by their interfaces, *in addition* to any thermal resistance of poor substrates (e.g. flexible plastics[5] or multi-layered "high-rise" processors)[3]. The TBC of $MoS_2$-$SiO_2$ is also two to four times lower than that of graphene-$SiO_2$ interfaces,[28] which is consistent with the four times heavier mass per unit area of $MoS_2$ compared to graphene.[26] Similar TBC values are expected for other 2D atomically thin layers (on $SiO_2$), the lowest potentially belonging to $WTe_2$, which has twice the mass density per unit area of $MoS_2$.

Before concluding, we note that our investigation also sheds light on the breakdown (BD) mechanism of such 2D devices. Figure 4a shows the temperature along the $MoS_2$ channel at the onset of breakdown, illustrating a hot spot forming near the drain. The device failed after being held at a lateral field $E \approx 5$ V/μm and current $I_D \approx 210$ μA/μm for several minutes. The AFM and scanning electron microscopy (SEM) images post-breakdown (Figures 4b,c) confirm the failure location. More than 20 devices were examined and all showed similar damage location after breakdown. Although the temperature measured by Raman is averaged across the spot size, the localized temperature can exceed the $MoS_2$ oxidation threshold $T_{BD} \approx 380$ °C when the power density is highly peaked at the drain. This behavior reveals that the ~10 nm thin $AlO_x$ capping layer used here is a good oxygen barrier at room temperature[21] (stabilizing the device during Raman measurements vs. uncapped devices, see Methods), but not at the elevated temperatures near $MoS_2$ breakdown. (See Supporting Information Section 10 for capped and uncapped $MoS_2$ oxidation studies, and Section 8 for thermal modeling.)

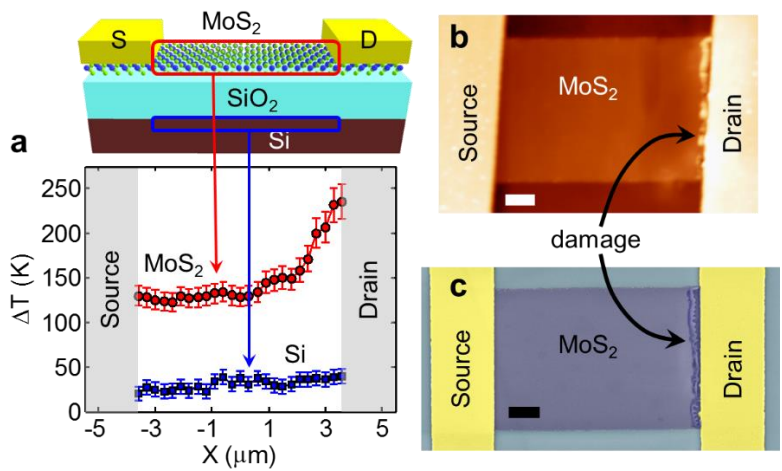

**Figure 4 | Device thermal breakdown. (a)** Temperature rise measured by Raman thermometry of $MoS_2$ (red) and Si substrate (blue) along the channel of a transistor operated near the onset of breakdown ($P \geq 1.05$ mW/$\mu m^2$). At high bias, heating becomes more pronounced at the drain side of the channel. This is attributed to a combination of channel pinch-off[29] and current crowding[30] at the drain contact. **(b)** AFM topography and **(c)** colored SEM of the $MoS_2$ channel post-breakdown showing damage near the drain contact, at the location of maximum device temperature. The scale bars are 1 $\mu m$.

In summary, we investigated energy dissipation in functioning monolayer $MoS_2$ transistors for the first time. Raman thermometry takes advantage of material selectivity, simultaneously measuring the temperature of the transistor and substrate. We uncover relatively uniform heating, even near small bilayer regions present in some CVD grown films, revealing that 2D semiconductors are more immune to such variability than expected. However, thermal breakdown occurs at the drain of such devices, when the (localized) temperature exceeds the oxidation threshold of $MoS_2$. We find that the $MoS_2$ interface will ultimately limit energy dissipation, and its TBC is among the lowest presently known for solid-solid interfaces. Such 2D electronics can nonetheless benefit from better thermal substrates (e.g. thinner $SiO_2$), while poor thermal substrates like flexible plastics could severely limit their performance.[20] Partial device cooling could be obtained from capping layers with higher thermal conductivity (e.g. h-BN), used in short-channel devices (<100 nm) where partial heat sinking can occur directly to the contacts[21] (Supporting Information Section 8). Overall, our findings shed new light on energy dissipation mechanisms in 2D semiconductor devices, paving the way towards energy-aware design of 2D electronics.

## Methods

CVD growth of monolayer $MoS_2$ was performed directly on $SiO_2$ (90 nm) on Si ($p^+$, electrical resistivity of 1-5 $m\Omega \cdot cm$) substrates at 850 ºC and 760 Torr with the aid of a PTAS seed layer to encourage large-grain epitaxial growth.[11] Small bilayer regions (<10% areal coverage) can form due to the size of the resultant grains being larger than the surface diffusion length. For comparison, exfoliated monolayer $MoS_2$ flakes were also prepared onto identical substrates (see Supporting Information Section 6).

Electron-beam (e-beam) lithography was used to define contact regions and channel dimensions (widths $W$ = 4.5–5 µm and lengths $L$ = 2.5–6.8 µm) for all $MoS_2$ devices. $MoS_2$ was etched using $XeF_2$ gas, followed by e-beam evaporation of 40 nm Au at low pressure (~$8 \times 10^{-8}$ Torr) and lift-off, to obtain ultra-clean Au contacts.[7] Underneath the probing pads a 4 nm Ti adhesion layer was deposited prior to 40 nm Au. All devices were annealed in vacuum (~$10^{-5}$ Torr) at 250 °C for 1 hour to improve contacts and remove surface adsorbates, then encapsulated to enable stable operation during extended thermal testing in ambient air. The capping layer consisted of e-beam evaporated and oxidized 1.5 nm Al seed, followed by atomic layer deposition (ALD) of ~15 nm amorphous $AlO_x$ by alternating[21] trimethylaluminum (TMA) and $H_2O$ pulses at 150 °C. The $AlO_x$ capping induced n-type doping of the $MoS_2$ channel.[31]

Raman spectroscopy was carried out using a Horiba LabRam instrument with a 532 nm laser and 100× long working distance objective with N.A. = 0.6. Step sizes in the Raman maps varied between 0.1–0.2 µm and the acquisition time of each device thermal map was ~10–15 minutes. The laser spot radius is ~0.3 µm, and the absorbed laser power is < 20 µW to avoid laser heating in excess of the electrical heating (see Supporting Information Section 5) and to maintain negligible photocurrent. Temperature calibration was done with a Linkam THMS600 stage. We corrected for artifact Raman shifts due to sample drift during the measurement (Supporting Information Section 4). Smaller shifts of the $A_1$' mode due to carrier density gradients along the channel[32] are also corrected in our analysis (Supporting Information Section 3).

All thermometry measurements were performed in air, at ambient temperature $T_0 \approx 20$ ºC. Electrical measurements were carried out using a Keithley 4200 and home-built probe station. Atomic Force Microscopy (AFM) images were taken with a Veeco® AFM system, and Scanning Thermal Microscopy (SThM) images were obtained using a commercial module from Anasys®.

## Acknowledgements

We thank Özgür Burak Aslan and Zhun-Yong Ong for helpful discussions, and acknowledge the Stanford Nanofabrication Facility (SNF) and Stanford Nano Shared Facilities (SNSF) for enabling device fabrication and measurements. This work was supported in part by the National Science Foundation (NSF) EFRI 2-DARE grant 1542883, the NSF DMREF grant 1534279, the Air Force Office of Scientific Research (AFOSR) grant FA9550-14-1-0251, the NCN-NEEDS program, which is funded by the NSF contract 1227020-EEC and by the Semiconductor Research Corporation (SRC), and in part by the Stanford SystemX Alliance. E.Y. acknowledges partial support from the Ilan Ramon Fulbright Fellowship and from the Andrew and Erna Finci Viterbi Foundation. K.K.H.S. and C.J.M. acknowledge partial support from the NSF Graduate Research Fellowship.

# Energy Dissipation in Monolayer MoS₂ Electronics

Eilam Yalon[1], Connor J. McClellan[1], Kirby K. H. Smithe[1], Miguel Muñoz Rojo[1], Runjie (Lily) Xu[1], Saurabh V. Suryavanshi[1], Alex J. Gabourie[1], Christopher M. Neumann[1], Feng Xiong[1,2], Amir B. Farimani[3], and Eric Pop[1,4,5,*]

[1]*Department of Electrical Engineering, Stanford University, Stanford, CA 94305, USA.* [2]*Present address: Department of Electrical & Computer Engineering, University of Pittsburgh, Pittsburgh, PA 15261, USA.* [3]*Department of Chemistry, Stanford University, Stanford, CA 94305, USA* [4]*Department of Materials Science & Engineering, Stanford University, Stanford, CA 94305, USA.* [5]*Precourt Institute for Energy, Stanford University, Stanford, CA 94305, USA.* [*]*E-mail:* [epop@stanford.edu](epop@stanford.edu)

## Supporting Information

## Table of Contents

## S1.    Electrical contact resistance

The electrical contact resistance was evaluated by the transfer length method (TLM). Figure S1 shows the total resistance $R_{TOT}$ (normalized by width) vs. channel length. We extract contact resistance $R_C = 1.6 \pm 2.5$ k$\Omega \cdot \mu$m with the uncertainty reflecting 95% confidence intervals from a least-squares fit of the TLM plot. To err on the conservative side, we only set an upper bound as the goodness of the TLM fit is limited by our shortest channel length $L_{min} = 0.5$ $\mu$m. We therefore set an upper bound of $R_C \leq 4$ k$\Omega \cdot \mu$m to be used when estimating the fraction of power dissipated at the contacts (see Supporting Information Section 10). For the extraction of thermal boundary conductance (TBC) discussed in this work we only used transistors with $L > 4$ $\mu$m, for which $R_C < 0.1$ $R_{TOT}$. We also subtracted the power dissipated at the contacts ($2I^2R_C$) from the total power input of all data in the main text Figure 3a.

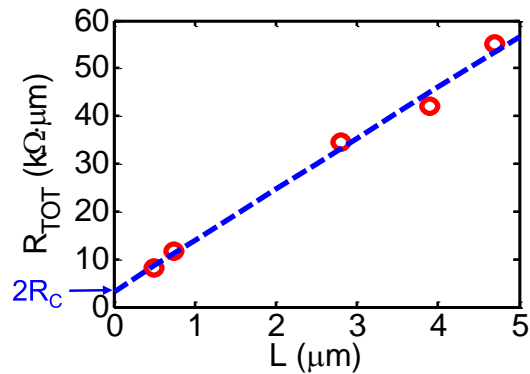

**Figure S1 | Electrical contact resistance.** TLM plot showing total resistance vs. channel length at high gate bias ($V_{GS} = 25$ V). The linear fit yields sheet and contact resistance. Additional details and uncertainty analysis follow discussion in Ref. 1.

## S2.    Temperature maps of exfoliated 1L MoS$_2$ devices

We compared our measurements of 1L CVD MoS$_2$ transistors to similar devices fabricated from exfoliated 1L MoS$_2$ channels. The exfoliated monolayer MoS$_2$ flakes were prepared using a gold-assisted exfoliation method[2] onto identical substrates as the CVD-grown devices. The exfoliated devices were also capped by ~15 nm AlO$_x$ (see Methods), being expected to be similarly doped as the CVD-grown devices. The obtained temperature distribution of the exfoliated devices from Raman spectroscopy (Figure S2) is uniform and the thermal resistance is comparable to the one obtained for CVD MoS$_2$ devices. (See Figure 3 of main text for multiple comparisons.)

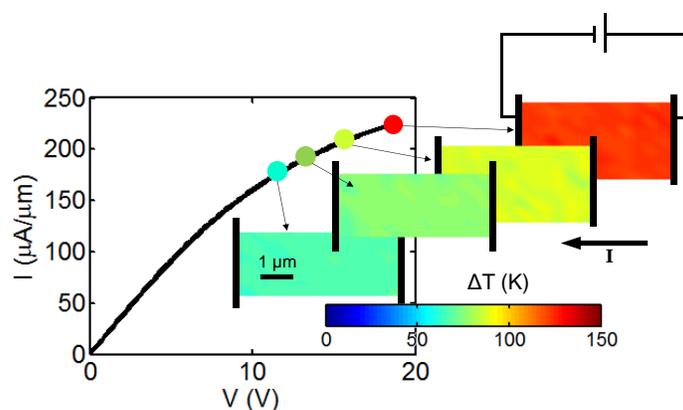

**Figure S2 | Temperature rise in exfoliated 1L MoS₂ devices.** Output characteristics of 1L MoS$_2$ transistor and $\Delta T$ maps at four different bias conditions. Filled circles mark the applied voltage and current of the respective temperature color maps. Measurement configuration is shown schematically on the top temperature map. Current flow direction is from right (electron drain) to left (electron source) in all maps as indicated by the arrow. The temperature is largely uniform.

## S3.    Non-temperature related Raman peak shifts

The temperature in our experiment is measured by monitoring the softening of the Raman modes, and it is therefore important to account for any Raman shifts not induced directly by temperature, such as strain and doping. We have used the $A_1'$ mode in our measurements to avoid the uncertainty in the Raman shift due to strain present in the $E'$ mode during the temperature calibration. We also calibrated Raman peak shifts of the $A_1'$ mode vs. temperature for 1L and 2L (bilayer) separately, and found that the Raman peak shift of the $A_{1g}$ mode vs. temperature of 2L is 0.015 $\pm$ 0.002 cm$^{-1}$/C (not shown here) and is very close to the $A_1'$ mode obtained from 1L.

In addition, the $A_1'$ mode peak position has a slight dependence on carrier concentration[3]. We decoupled the carrier concentration dependence (induced by back-gate voltage, $V_{GS}$) from the temperature dependence in our measurement by calibrating the peak shift vs. $V_{GS}$ at $V_{DS}$=0 as shown in Figure S3. We then corrected the Raman signal across the device length by subtracting the peak shift induced by $V_{GS}$ -$V_{(x)}$, where $V_{(x)}$ is the voltage at position x in the channel, assuming linear voltage distribution between source (x = −L/2) to drain (x = L/2).

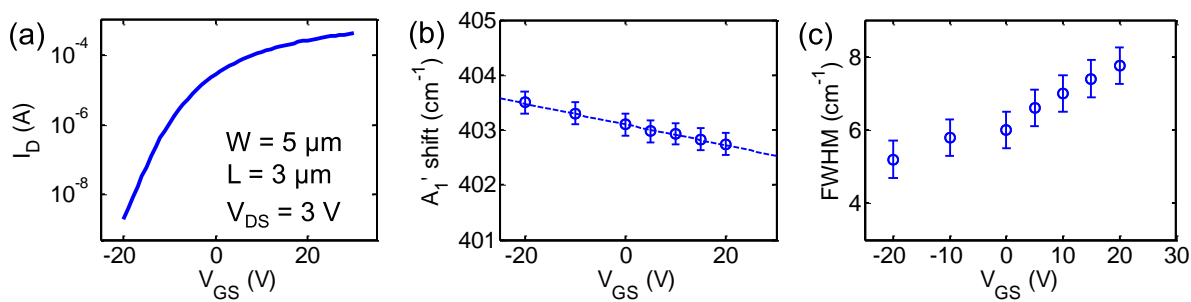

**Figure S3 | Gate voltage dependence of Raman $A_1'$ mode.** (a) $I_D$ vs. $V_{GS}$ of a representative device. (b) Raman peak shift and (c) full-width at half-maximum (FWHM) broadening of the $A_1'$ mode vs. $V_{GS}$. The $E'$ mode did not show changes in peak position or FWHM.

## S4.    Corrections for stage drift

We obtain the spatially resolved temperature by Raman mapping of our devices with and without electrical bias, and comparing the Raman peak shifts to their temperature calibration done on a hot stage. Since the peak position of $MoS_2$ out-of-plane Raman mode ($A_1$' in 1L and $A_{1g}$ in 2L) depends on the number of layers, the Raman signal in the presence of small 2L regions is non-uniform. The 2L $A_{1g}$ mode in our samples is higher by ~2 cm$^{-1}$ compared with the 1L $A_1$' mode, in agreement with previous reports for the same laser wavelength. During data analysis, this non-uniform Raman signal across the device, induced by the presence of 2L regions, must be carefully examined. In addition, small shifts in the sample position (~100 nm) result in misalignment between the reference and the biased Raman maps and must be corrected.

We present our correction method in Figure S4 by comparing a reference map (a) acquired at room temperature and a "hot" map (b) acquired at stage temperature $T_{stage} = 175$ ºC. In this case the device temperature should be uniform as no bias is applied. Figure S4c shows the raw temperature extraction obtained directly by subtracting map (b) from (a) and dividing by the calibration value (Raman peak shift to temperature) from Figure S5d. It is evident that the extracted *raw* ΔT map is non-uniform in temperature and includes artificially hot and cold spots. These artificial non-uniformities in temperature can be associated with the drift of the stage during the measurement. We note that the typical acquisition time of these Raman maps is of the order ~10 minutes, and even drift of ~150 nm is sufficient to induce the observed changes.

We have therefore developed a correction procedure that includes dividing the map into areas and sorting the spectra of different pixels by their Raman peak position (or intensity). We then subtract the pixels of each area one by one in their order (as they were sorted), such that the pixel with the highest Raman signal of one area is aligned with the pixel of the highest Raman signal in the same area of the reference map. We assume the temperature does not shift one pixel significantly more than the other such that, for example, a 2L pixel having its Raman peak 1 cm$^{-1}$ higher than a 1L pixel at room temperature will not shift to a lower wavenumber than the 1L pixel at high temperature. The reason is that the difference in $A_1$' peak position from 1L to $A_{1g}$ in 2L (~2 cm$^{-1}$) is large compared with any possible temperature variations across the sample. The uniform temperature map (within the uncertainty of the Raman measurement) in Figure S4d confirms our correction procedure since the temperature across the device is expected to be uniform when heated on a hot stage (rather than heated by electrical bias). We note that this calibration procedure is necessary only when the Raman signal across the measured area is non-uniform.

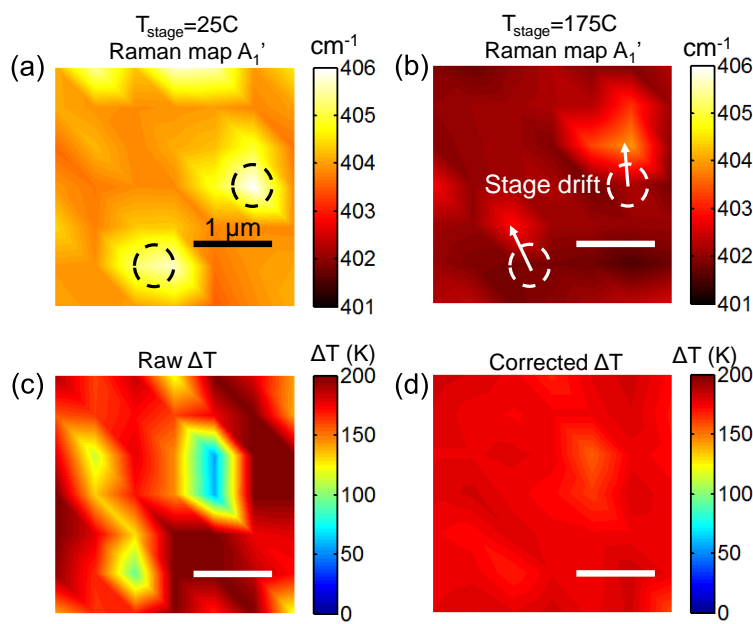

**Figure S4 | Correction of non-uniformity and misaligned Raman maps for temperature extraction.** (a) Raman $A_1'$ peak position mapping at room temperature showing non-uniformity due to small 2L spots (dashed circles). (b) Raman $A_1'$ peak position mapping at $T_{stage} = 175$ °C showing softening of the $A_1'$ mode to lower wavenumbers and (non-uniform) stage drift visible by the change in the location of the 2L spots. (c) Raw temperature map obtained by subtracting Raman map (b) from (a) and dividing by the calibration of peak shift vs. temperature from Figure S5d. The map shows "artificially" hot and cold spots due to the misalignment and subtraction of 2L (1L) from 1L (2L) Raman signal. (d) Temperature map after applying the procedure outlined in the text correcting for misalignment of the non-uniform Raman maps.

## S5.    Temperature-dependent Raman spectroscopy of monolayer MoS$_2$

Raman temperature measurements were carried out by comparing the shifts in spectral peak position under applied bias (with respect to the unbiased case) to a calibration measurement on a hot stage, where the sample temperature was known. We note that the Stokes to anti-Stokes intensity ratio can also be used as a thermometer[4, 5], however it relies on the measurement of intensity rather than spectral peak position; the latter being more accurate in our measurements. The Stokes to anti-Stokes ratio is also not suitable for measuring temperature when the incident laser energy lies close to an excitonic state energy and resonance effects dominate the measured intensity, as was the case in this study.

The calibration of Raman peak shift with temperature was carried out in five different locations on films similar to the ones measured electrically up to 250 °C – monolayer (1L) CVD and exfoliated MoS$_2$ capped by AlO$_x$. Data from a representative location is shown in Figure S5. In addition, we carried out the same procedure on the MoS$_2$ transistors that were measured electrically, but only up to 125 °C in order not to degrade their performance. We found that the absolute peak position slightly varied between samples, however the peak shift with temperature (the slope in Figure S5) was similar across different locations and different samples, within the uncertainty of the measurement (error-bars in Figure S5). The absorbed laser power here and in the electrical measurement is kept below 20 µW, such that the temperature rise induced by the laser is always < 8 °C. This is confirmed by the observation that the Raman modes do not shift within the uncertainty of the measurement between 1.5 µW and 20 µW incident laser power. For the Si substrate, the absorption depth of the 532 nm laser in highly doped Si is ~ 0.65 µm.[6] Given the dimensions of the device ($4 \times 5$ µm$^2$) and the Si substrate thickness (500 µm) we can consider the measured temperature as that of the Si surface.

The temperature dependent Raman spectra in our devices agree with previous reports of 1L MoS$_2$ on SiO$_2$[7, 8]. The temperature dependence of the out-of-plane A$_1$' mode was consistent in all measured devices, whether the MoS$_2$ was grown by CVD or exfoliated, and capped with AlO$_x$ as well as uncapped. The in-plane E' mode showed some variations between different types of samples, possibly due to strain (e.g. grown by CVD vs. exfoliated). In addition we note that for MoS$_2$ grown on quartz we measured the E' mode peak at higher frequency (~1.5 cm$^{-1}$ higher than E' of MoS$_2$ on Si/SiO$_2$) whereas the A$_1$' mode maintained its peak position. Similarly, previous studies showed E' mode spectral response was different between MoS$_2$ on Si$_3$N$_4$ and sapphire substrates, whereas A$_1$' maintained its spectral response with both substrates[7]. We have therefore used the shifts in A$_1$' Raman mode as the thermometer in our measurements. The uncertainty in temperature measurement of the MoS$_2$ is 5-10 K (Figure 3a), whereas the uncertainty in the temperature measurement of the Si is about half, since the sensitivity of its Raman shift to temperature is almost double, as evident in Figure S5.

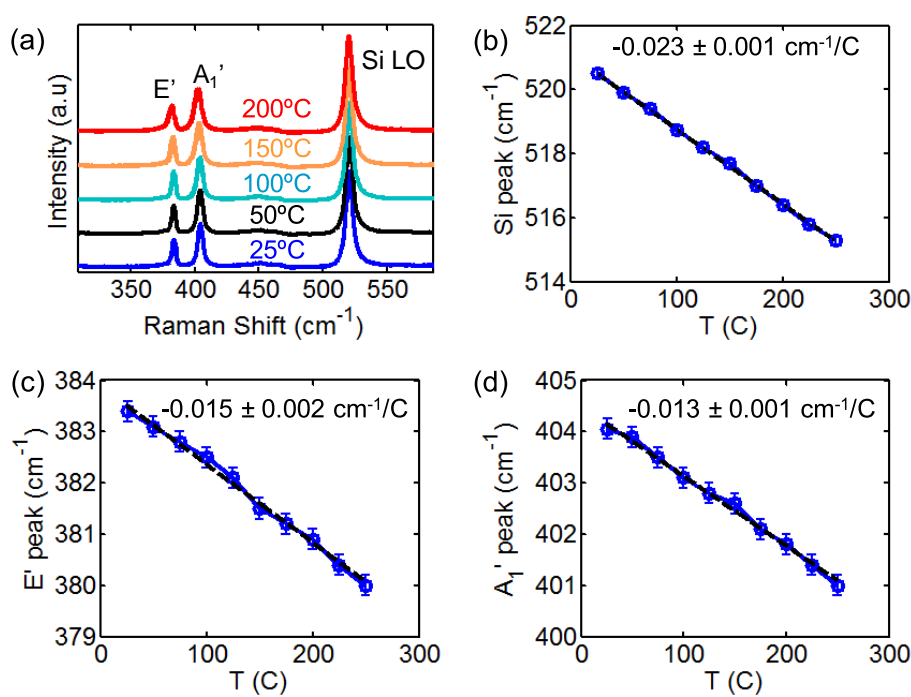

**Figure S5 | Temperature-dependent Raman spectra.** (a) Raman spectra of CVD 1L MoS$_2$ (on 90 nm SiO$_2$ on Si) at varying temperatures. Raman shift vs. temperature of (b) Si substrate zone-center longitudinal optical (LO) phonon, (c) MoS$_2$ channel E' and (d) A$_1$' phonon modes. The Si substrate LO phonon Raman shift is in good agreement with previous studies[9].

# S6.    Scanning thermal microscopy (SThM)

We confirmed the uniform distribution of temperature rise in our devices by scanning thermal microscopy (SThM) measurements, as follows. A commercial SThM module from Anasys® Instruments was added onto an Atomic Force Microscope (AFM) from Veeco® Instruments. SThM usually consists of a thermo-resistive probe that is connected to a Wheatstone bridge, a DC voltage source and an amplifier specifically designed to avoid small electrical spikes that could break the probe. Temperature sensing occurs when the sample (here the $AlO_x$ capping layer covering the $MoS_2$ transistor) heats up, and the SThM tip changes its electrical resistance. Using this technique, a thermal map of the sample surface with nanoscale resolution can be obtained[10]. The thermal probe used in this work is DM-GLA-5 provided by Anasys®, made of a thin Pd layer on SiN.

The $MoS_2$ device was placed on the AFM holder, and its electrical pads were wirebonded to small pieces of Au on $SiO_2$/Si substrates with areas of ~0.5 x 0.5 $cm^2$ and total thickness of ~500 μm. Thin copper wires with radius ~50 μm were contacted to these substrates using silver epoxy. These wires were used to apply current through the $MoS_2$ film using an electrical source. The $MoS_2$ transistor is capped with 15 nm of $AlO_x$, which prevents the SThM probe from electrical discharges that could break the probe, but results in measurement of the top $AlO_x$ surface rather than direct measurement of the $MoS_2$ channel.

We heated the transistor electrically by applying voltage to the $MoS_2$ channel, while using the SThM probe to obtain a thermal map of the device. The measured signal is proportional to temperature but is qualitative. The temperature scale-bar used in Figure S6 is estimated from our thermal model calibrated by Raman thermometry. We note, however that the temperature resolution of the SThM measurement (<5 K) is better than that of Raman (~10 K). The SThM detects temperature rise at low input power for which $\Delta T$ is lower than the uncertainty of the Raman measurement, confirming the higher temperature sensitivity of the SThM.

The high spatial resolution of SThM confirms that the small 2L regions of $MoS_2$ do not act as hot spots. Another interesting feature observed in the SThM maps is some cooling at the $MoS_2$ channel edges. The small asymmetry in the two gradients observed at both sides of the film edges can be considered as a probe artifact. The slightly larger gradient on one of the sides happens when the probe lifts from the $SiO_2$ to the $MoS_2$ film, which causes some instability in the thermal scan, while the gradient observed on the other edge of the film is better represented, since the probe goes from $MoS_2$ film to the $SiO_2$. The decrease in temperature at the edges of the $MoS_2$ film is also found in finite element thermal simulations shown in Figure S10c.

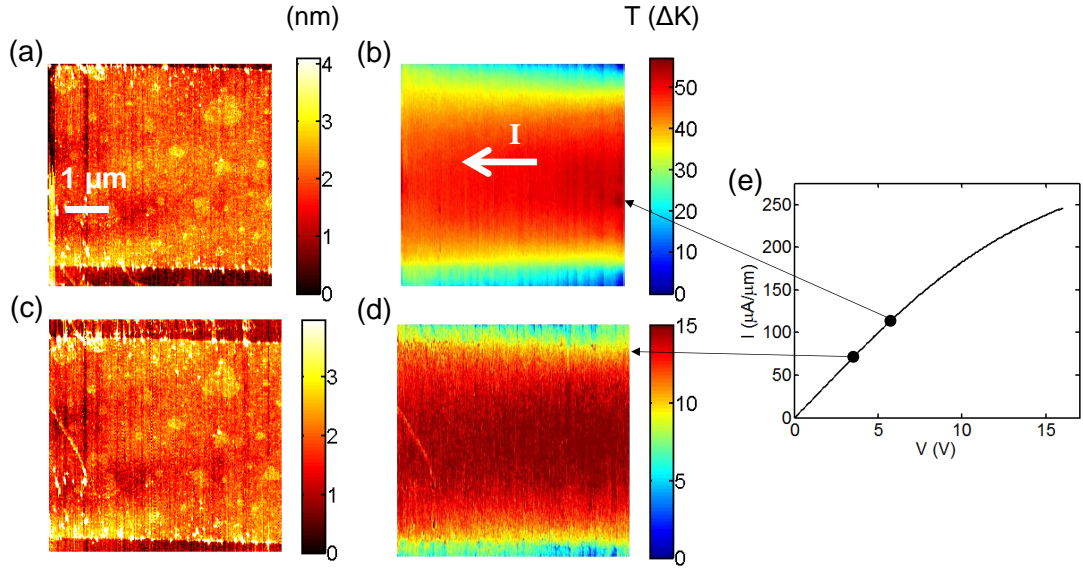

**Figure S6 | AFM and SThM.** (a,c) AFM topography maps (b,d) SThM maps of the same MoS$_2$ channel from Figure 2 of the main text, with electrical bias shown in (e). The AFM and SThM are acquired simultaneously (a with b; c with d). AFM images show small 2L MoS$_2$ regions in the channel, and nucleation of the AlO$_x$ capping at the edges. SThM shows the temperature rise is uniform, with some cooling at the edges as confirmed by simulations in Figure S10c.

## S7.  Temperature estimates at 1L-2L junctions

The uniform heating observed in our CVD MoS$_2$ channels which include some 2L features implies that heating at 1L-2L junctions is smaller than the uncertainty in the Raman and SThM temperature measurements. This finding allows us to estimate: 1) the maximum electrical resistance of the 2L-1L junctions (R$_{2L-1L}$), and 2) the maximum conduction band (CB) offset between 1L and 2L MoS$_2$. The former results in Joule heating when electrons cross the $\Delta E_{CB}$ barrier from 2L to 1L, the latter results in thermionic heating when hot electrons dissipate their energy after injection from 1L to 2L. (The CB of 1L MoS$_2$ is nominally expected to be ~50 meV higher than for 2L[11].)

In Figure S7, we carried out thermal simulations of our device with channel length L = 4 μm with uniform power density, and placed an additional power generation source at the center of the channel to simulate (possible) additional heating at a 1L-2L junction. We set the length of the additional heat source to be of the order of the electron mean free path (λ$_{MFP}$ ≈ 2 nm, see Figure S10 in Supplement of Ref. 12). We varied the power density of the junction heat source to find the conditions that would have resulted in measurable heating. Figure S7a shows the temperature rise along the channel for different power densities at the junction (P is the uniform power density in the channel). Figure S7b shows the temperature rise in the channel when the heat source at the junction is set to 20P along with the temperature that would be measured by Raman (Gaussian average across the laser spot size) and SThM (due to heat spread in the capping layer and thermal exchange radius of the tip).[13] We find that in order to detect over-heating at the junction by Raman thermometry and SThM the power density at the junction must be: 1) higher than P$_{min}$ = 20 μW/μm, and 2) higher than 20P, where P is the (uniform) power density in the rest of the channel. The former condition is derived from a similar plot to Figure S7a but with the background temperature rise of $\Delta T \approx 0$ (not shown here). Since over-heating at the junction is not observed experimentally by Raman and SThM, we can estimate the power dissipated at the junction is smaller than the conditions outlined above. We note that 1L-2L and 2L-1L junctions could lead to either thermionic heating or cooling (depending on current flow direction), and neither effect is detectable here.

In Figure S7c we derive the minimum CB offset that would result in measurable heating at the junction based on these two conditions. The minimum power density curve is shown in blue, the 20P curve is shown in black and the red curve satisfies both conditions. We use an Ohmic current-voltage relation and assume uniform electric field (E) distribution in the channel, such that $P = V^2/R = E^2L^2/R$, where the channel lengths (L = 2.5 to 6.8 μm) and sheet resistance (R ~ 13 kΩ/□) are obtained from our measured devices. We assume the power dissipated by hot electron injection at the junction is $P_{1L-2L} = \Delta E_{CB}I$ which determines the condition $P_{1L-2L} = 20P$ as $\Delta E_{CB} = 20E\lambda_{MFP}$ (black dashed line in Figure S7c).

The minimum CB offset required to induce measurable heating is found to be ~120 meV. Since no over-heating was detected we conclude that the CB offset between 1L and 2L in our devices is smaller than $\Delta E_{CB} < 120$ meV. This finding agrees with recent experimental reports on the surface potential difference between 1L and 2L of the order of ~50 meV[14]. Similarly, one can estimate based on the same power dissipation requirements ($R_{2L-1L} < 20R\lambda_{MFP}$) the maximum electrical resistance of the junction between 2L and 1L $MoS_2$ is $R_{2L-1L} < 500$ Ωμm.

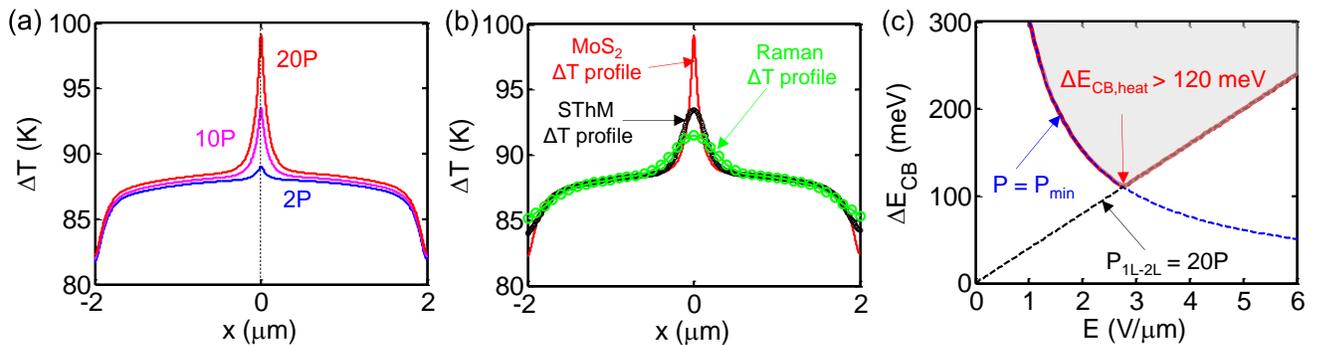

**Figure S7 | Thermal simulation of power dissipation at a line defect.** (a) Simulated temperature rise of 4 μm long monolayer $MoS_2$ channel with uniform power density P and an additional heat source (from 2P to 20P) at a line defect, such as a 1L-2L junction. (b) Same as (a) with 20P at the center (red line) and the temperature that would be measured by Raman (green circles, Gaussian average across the laser spot size) and SThM (black circles, representing the heat spread in the capping layer and the thermal exchange radius of the SThM tip).[13] (c) Maximum conduction band offset ($\Delta E_{CB}$) between 1L and 2L $MoS_2$ vs. E-field illustrating the regime for which measurable heating would be generated at a line defect such as a 1L-2L junction. Black dashed line satisfies $P_{1L-2L} = 20P$, blue dashed line satisfies $P_{1L-2L} = P_{min} (= 20$ μW/μm) and the red curve satisfies both. Heating would be measurable in the area shaded gray. Since no heating was measured at the junction, we estimate $\Delta E_{CB} < 120$ meV at 1L-2L junctions.

## S8.    Thermal analysis and modeling

We used the analytical model reported in Ref. 15 to extract the TBC from the measurements shown in Figure 3 of the main text. In the model we have used the Si thermal conductivity extracted from the slope of Si temperature vs. power density (~95 $Wm^{-1}K^{-1}$ which agrees well with known values for highly doped Si[16]). We also used known thermal conductivity of thermally-grown $SiO_2$ (1.4 W/m/K) and of the Si-$SiO_2$ TBC (> 125 $MWm^{-2}K^{-1}$).[17, 18] We note that the role of the Si-$SiO_2$ TBC here is negligible (see Figure S10d), accounting for < 5% of the total thermal resistance, but it could play a greater role in devices on thinner oxides (e.g. < ~ 25 nm). This is evident in Fig. 3 of Ref. 19 where a measurable effect of the TBC is only observed for $SiO_2$ thinner than 25 nm.

We approximate the expression for the spreading thermal resistance to the Si substrate in Figure 3c of the main text with the shape factor of a circular disk heater on a semi-infinite substrate. To test the validity of this expression we carried out finite element thermal simulations of the structure used in this study (rectangular heater W × L = 5 × 4 $\mu m^2$). We found that the circular disk expression is within less than 5% error of the numerically accurate thermal resistance for the average temperature of a rectangular heat source. Figure S8 shows the temperature distribution of the thermal spreading to the Si substrate in the finite element simulation illustrating the circular profile of the temperature.

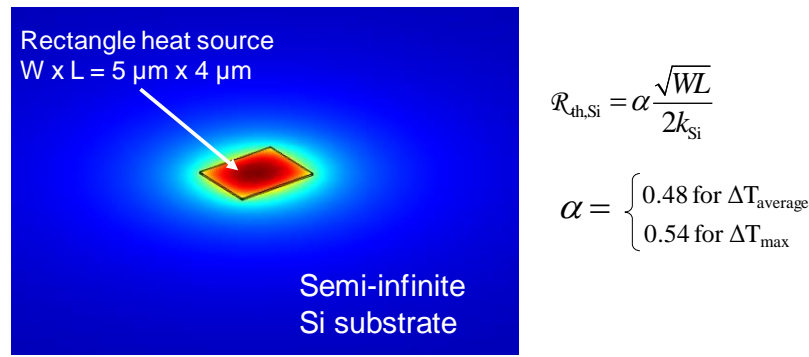

**Figure S8 | Simulated thermal spreading to the Si substrate.** Simulated temperature distribution for the case of rectangular heater (W × L = 5 × 4 $\mu m^2$) and thermal spreading to the Si substrate. The temperature profile is circular and the approximation of circular disk shape heater on semi-infinite substrate is good within less than 5% error.

The $MoS_2$-$SiO_2$ TBC is extracted by subtracting the Si and $SiO_2$ thermal resistance contribution to the total thermal resistance $\mathcal{R}_{th} = \Delta T/P$, as in Figure 3 of the main text. Here P is the power density in the $MoS_2$ channel after the contact power dissipation ($2I^2R_C$) was subtracted, as stated in Supporting Information Section 1.

Figure S9 presents a histogram of all extracted TBC values. The histogram is fitted to a normal distribution with mean (± standard deviation) of 14 ± 4 $MWm^{-2}K^{-1}$. Variations in our TBC values are due to uncertainties in contact resistance, Raman shift measurement, and peak fitting.

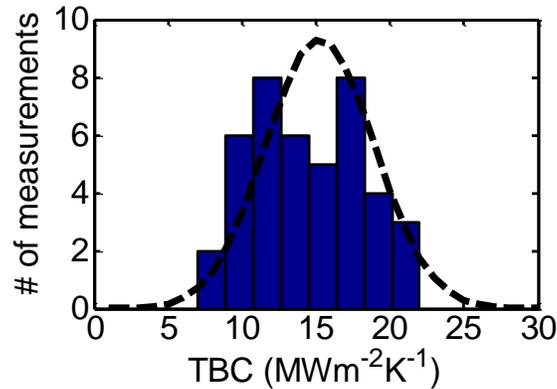

**Figure S9 | Measured MoS₂-SiO₂ thermal boundary conductance (TBC) distribution.** Histogram of all (CVD and exfoliated) MoS₂-SiO₂ TBCs extracted from the measurements shown in Figure 3a of the main text. More than 40 measurements of 8 CVD and 3 exfoliated devices (at varying input power) are shown. Dashed black line represents normal distribution with mean (± standard deviation) of $14 \pm 4$ MWm$^{-2}$K$^{-1}$.

We note that the doping induced by the AlO$_x$ capping layer prevents pinch-off and results in uniform heating in the channel, except at the onset of breakdown when heating becomes more significant at the drain (Figure 4 of main text). The measured *uniform* temperature rise justifies the use of the analytic model, whereas a non-uniform power dissipation model should be invoked at the onset of breakdown. We also compare our experimental results with finite element electro-thermal simulations (COMSOL Multiphysics software ®) to confirm the analytic model. The simulation results are summarized in Figs. S10-S12. The voltage drop and consequent heat generation at the contacts are included in the electrothermal simulations, yet most of the power (>90%) is dissipated at the channel as indicated in Supporting Section 1.

The lateral temperature distribution shows some cooling to the contacts (along the channel) and sideways (across channel width) within a characteristic thermal healing length[20] L$_H$ ~ 100 nm. The AlO$_x$ capping adds a parallel path for lateral heat flow to the contacts, hence increasing L$_H$ compared to the uncapped devices (Figure S11). The temperature decay sideways (across channel width) within a length scale L$_H$ is qualitatively captured by the SThM (Figure S6), but since L$_H$ < laser spot size, the effect is not captured by the Raman measurements.

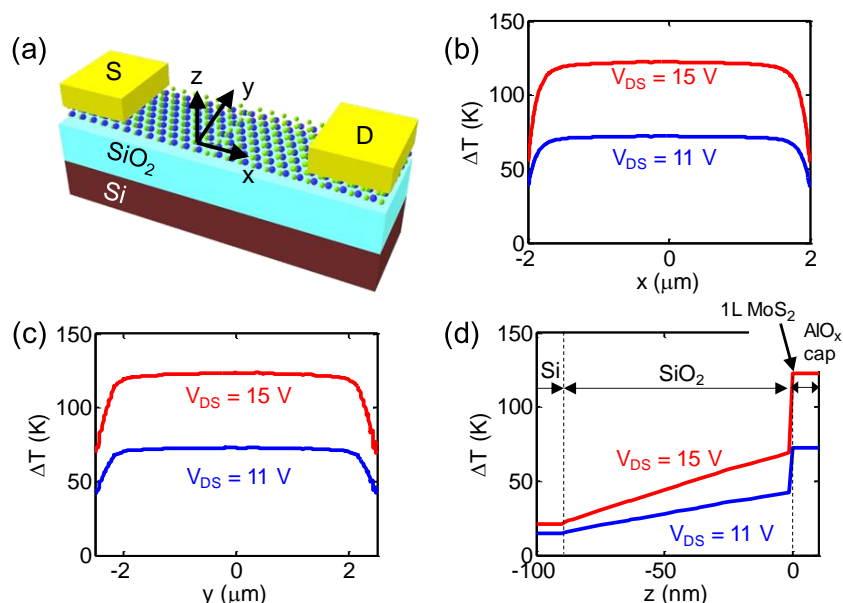

**Figure S10 | Electro-thermal simulations.** (a) Simulated structure showing the x (length), y (width) and z (height) axis. Simulated steady-state temperature rise at two applied electrical biases along: (b) device length, (c) width, and (d) vertical axis. Device dimensions L = 4 µm, W = 5 µm. (b) and (c) show some cooling to the contacts and sideways within a characteristic thermal healing length $L_H \sim 100$ nm. The vertical z-axis (d) shows that the capping layer is at the same temperature as the $MoS_2$ channel, a large $\Delta T$ across the $MoS_2$-$SiO_2$ interface due to its thermal boundary resistance, gradual T decrease in the $SiO_2$ ($k_{SiO2} \sim 1.4$ $Wm^{-1}K^{-1}$), a negligible $\Delta T$ at the $Si/SiO_2$ interface, and gradual decrease in temperature into the Si substrate ($k_{Si,doped} \sim 95$ $Wm^{-1}K^{-1}$).

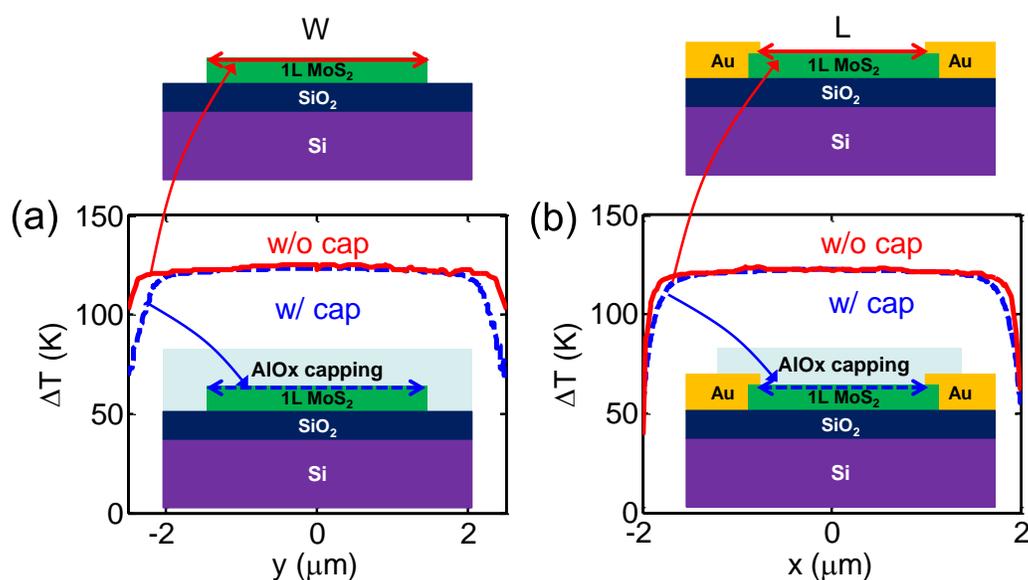

**Figure S11 | The thermal role of a capping layer.** Simulated steady-state temperature rise across (a) channel width and (b) length with (blue) and without (red) a capping layer of 15 nm $AlO_x$. Axis (y for channel width and x for channel length) as defined in Figure S10a. The temperature distribution with a capping layer is very similar to the one without it, except the thermal healing length is slightly longer since some of the heat is carried laterally by the capping layer.

Finally we note that for devices used in this study, where W and L $\gg$ $L_H$ and no top-gate is present, the simplified lumped model presented in Figure 3c of the main text can readily be used.

We also illustrate via thermal simulations how the peak device temperature at nanoscale hot spots near the drain could be higher than the one measured by Raman. The temperature measured by Raman follows a Gaussian weighed function with laser beam size $r_0$:[21]

$$T_{Raman} = \frac{\int\limits_0^\infty T(r)\exp\left(-\frac{r^2}{r_0^2}\right)r\,dr}{\int\limits_0^\infty \exp\left(-\frac{r^2}{r_0^2}\right)r\,dr} \tag{2}$$

Figure S12 shows the simulated temperature profile along the $MoS_2$ channel and the temperature that would be measured by Raman with a beam size $r_0 \approx 300$ nm (measured experimentally in our devices by the knife edge method[21]). The simulated temperature was chosen to represent the onset of thermal breakdown, having a peaked profile with a ~20 nm hot spot at the drain exceeding the oxidation temperature of $MoS_2$ ($T \approx 400$ °C > $T_{BD} \approx 380$ °C). The peak temperature "seen" by Raman thermometry is ~300 °C. The difference between the local temperature (on nm-scale) and the one measured by Raman can account for the difference between the maximum temperature measured in Figure 4a of the main text and the oxidation temperature of $AlO_x$-capped $MoS_2$ shown in Figure S15, required to initiate the thermal breakdown shown in Figs. 4b,c of the main text.

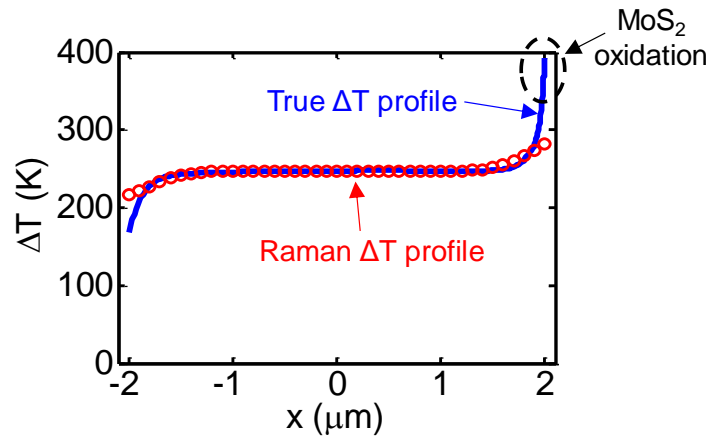

**Figure S12 | Predicted Raman measurement of a sharply peaked temperature profile.** Simulated temperature profile along $MoS_2$ channel illustrating hot spot (~20 nm) at drain side (blue) and the temperature that would be measured by Raman, following Eq. (2) with $r_0 = 300$ nm (red circles). The peak channel temperature at the drain exceeds the $MoS_2$ oxidation temperature ($T_{BD} \approx 380$ °C) but the maximum temperature measured by Raman is ~100 °C lower due to spatial averaging.

## S9.    Molecular dynamics (MD) simulations

To replicate the experimental setup within MD simulations, we use a simulation box containing a single layer of $MoS_2$ and $SiO_2$ as the substrate, as shown in Figure S13a. The substrate is a block of amorphous $SiO_2$ with dimensions $5.7 \times 5.7 \times 5.7$ nm created by the Visual Molecular Dynamic (VMD) package[22]. Periodic boundary conditions (PBC) were applied in all three directions. The x-y PBCs are chosen to create a continuous $MoS_2$ sheet. A vacuum region of 20 nm above the $MoS_2$ sheet is created to avoid interaction between the adjacent unit cells in the z-direction (perpendicular to the $MoS_2$ sheet). Initially, the distance between $MoS_2$ and $SiO_2$ is set to be at 3 Å.

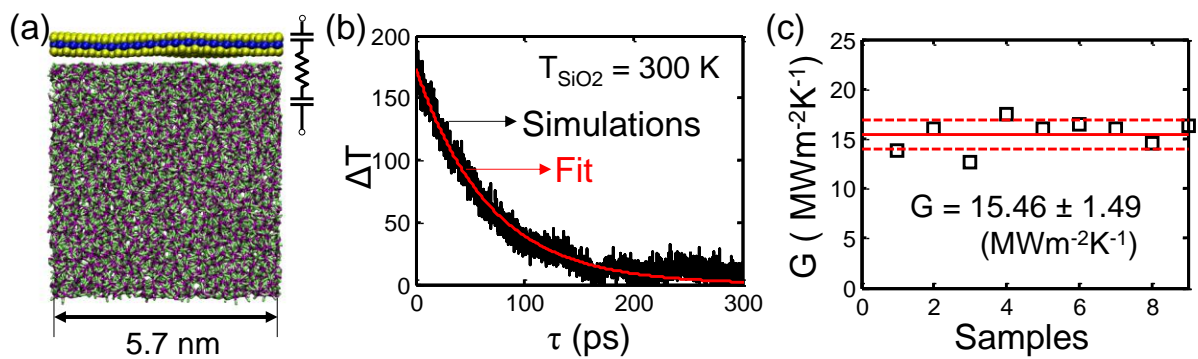

**Figure S13 | MD simulation of $MoS_2$ on $SiO_2$.** (a) 1L $MoS_2$ on small block of amorphous $SiO_2$. The inset shows the RC (resistance-capacitance) thermal circuit used to fit the exponential decay of $\Delta T$.  (b) Typical calculated decaying $\Delta T$ (black curve) and corresponding fit using eq. 1 (red curve). (c) TBC extracted from 9 different MD simulations. The solid red line is the mean (15.46 $MWm^{-2}K^{-1}$) and the dashed red lines show the estimated error (1.49 $MWm^{-2}K^{-1}$) from the 9 simulations.

We use the Tersoff potential[23] for $SiO_2$ and the Stillinger-Weber (SW) potential developed by Jiang *et al*. for $MoS_2$[24]. The interaction between $MoS_2$ and $SiO_2$ is modeled using Lorentz-Berthelot mixing rules. The (Lennard-Jones) LJ parameters for $MoS_2$ and $SiO_2$ were used based on universal force field[25] and are shown in Table S1.

|        | $\sigma$ (Å) | $\epsilon$ (meV) |
|--------|--------------|------------------|
| Mo-Si  | 3.27         | 6.52             |
| Mo-O   | 2.92         | 2.51             |
| S-Si   | 3.72         | 14.39            |
| S-O    | 3.38         | 5.56             |

**Table S1 |** LJ parameters ($\sigma$ and $\epsilon$) used for the interaction between 1L $MoS_2$ and $SiO_2$.

In order to stabilize the $SiO_2$ block, we first performed a separate equilibration simulation with $SiO_2$. This equilibration is performed in an NPT ensemble at the temperature of 300 K and constant pressure of 1 bar. The total simulation time for NPT was 200 ps with a time step of 0.01 fs. The small time step ensures the relaxation of $SiO_2$ atoms. In all simulations, we used Nosé-Hoover thermostat and Berendsen barostat to keep the temperature and the pressure constant. A single-layer of $MoS_2$ is then placed on the $SiO_2$ block. We performed the energy minimization of the system using the steepest decent algorithm. The tolerance for energy and force are both set at $10^{-6}$ and $10^{-6}$ eV/ Å respectively. We then perform a final equilibration step in an NPT ensemble (at T = 300 K and P = 1 bar) for 200 ps and with a time step of 0.01 fs.

To compute the TBC between $MoS_2$ and $SiO_2$, we set the temperature of $MoS_2$ and $SiO_2$ to 480 K and 300 K, respectively. This can be achieved by using separate thermostats for $MoS_2$ and $SiO_2$. At the set temperatures, the system is allowed to equilibrate for 1 ns in an NVT ensemble with a time step of 0.1 fs. After the temperatures of $MoS_2$ and $SiO_2$ reached equilibrium, we switch to an NVE ensemble where the energy of the whole system is conserved. We simulate in an NVE ensemble for 300 ps with a time step of 0.05 fs. As a result, the temperature of $MoS_2$ decreases while the temperature of $SiO_2$ increases slightly.

We calculate the difference in the temperature of the $MoS_2$ layer and the $SiO_2$ block ($\Delta T = T_{MoS2} - T_{SiO2}$) and fit it to an exponential decay as[26]:

$$\Delta T = \Delta T_0 e^{-\left[\frac{1}{m_{SiO2}C_{SiO2}} + \frac{1}{m_{MoS2}C_{MoS2}}\right]GA\tau}$$

(1)

where, $\Delta T_0$ is the initial temperature difference between $MoS_2$ and $SiO_2$ (here set to 180 K). The $m_{SiO2}$ and $m_{MoS2}$ are the masses of the $SiO_2$ block and the $MoS_2$ layer, respectively. The $C_{SiO2}$ and $C_{MoS2}$ are the specific heat per unit mass for the $SiO_2$ and $MoS_2$ respectively. $A$ is the total surface area between $MoS_2$ and $SiO_2$ and $\tau$ is the simulation time. The TBC is given by $G$.

In order to get sufficient statistics, we performed 9 simulations with different starting velocities, and the error bar is generated based on these samples. Finally, we obtain the TBC of $G = 15.46 \pm 1.49$ MWm$^{-2}$K$^{-1}$. These values are consistent with the experimental value of $14 \pm 4$ MWm$^{-2}$K$^{-1}$ as discussed in the main manuscript and in Section 10 below.

We note that the size of the $MoS_2$ is chosen large enough to get significant statistics and avoid the non-idealities that might be introduced due to extremely small unit cell.[27] We also performed the dependence of TBC on the thickness of $SiO_2$ and observe that the TBC does not change for $SiO_2$ thickness greater than 2 nm.

## S10. $MoS_2$ oxidation

We measured the $MoS_2$ oxidation temperature in air ambient with and without the $AlO_x$ capping layer in order to compare it to the thermal breakdown (BD) temperature of our devices in air, and to test the role of $AlO_x$ encapsulating the channel. We increased the stage temperature between 360 ºC and 420 ºC in 20 ºC increments and waited 10 minutes at each temperature. Optical images of the oxidation process are shown in Figure S14 for uncapped films and in Figure S15 for films capped with 15 nm $AlO_x$ deposited by ALD (see Methods).

Without a capping layer we observe $MoS_2$ oxidation (spots larger than few hundred nm) starts between 380 ºC and 400 ºC. At 420 ºC more than ~10 µm long spots are oxidized within minutes. The oxidation originates mainly from the centers of the $MoS_2$ triangles, ostensibly due to the presence of $MoO_x$ at these locations. The capped films oxidize at similar temperatures but at lower rates. At 420 ºC more than ~1 μm long spots are oxidized within minutes. Overall, the AlO$_x$ capping efficiently protected the MoS$_2$ during testing in ambient air, enabling stable device behavior, although it does not appear to prevent oxidation at high temperatures (~400 ºC). Therefore, the breakdown of our devices in air most likely occurs when the maximal local temperature (in the vicinity of the drain contact) reaches the oxidation temperature, $T_{BD} \approx 380$ ºC.

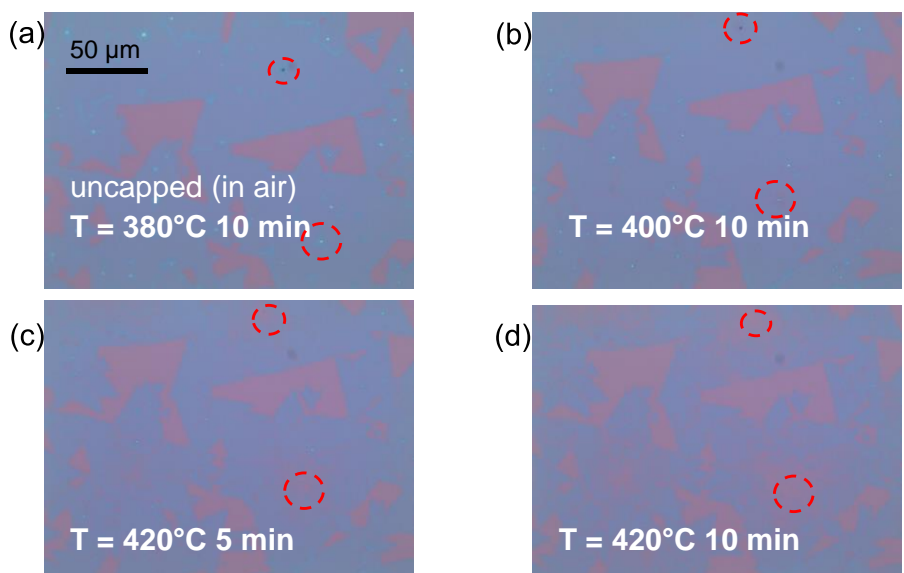

**Figure S14 | Oxidation of uncapped MoS$_2$.** Optical images of (uncapped) CVD MoS$_2$ after heating in air for (a) 10 minutes at 380 ºC, (b) 10 min. at 400 ºC, (c) 5 min. at 420 ºC, and (d) 10 min. at 420 ºC. Red dashed circles mark locations of oxidation, evidently initiating at nucleation centers, possibly due to presence of MoO$_X$. At 420 ºC after a few minutes, few tens of μm are oxidized.

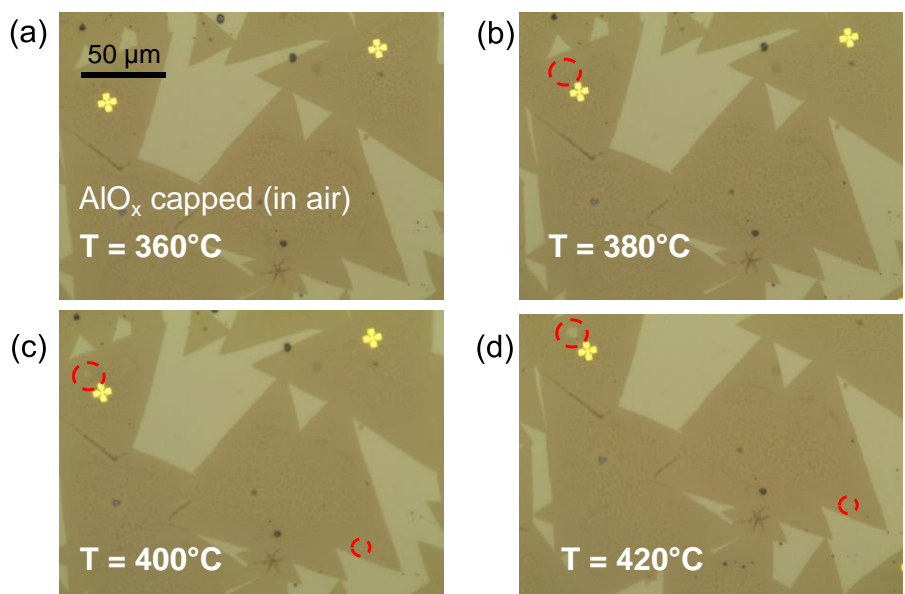

**Figure S15 | Oxidation of MoS$_2$ capped by AlO$_x$.** Optical images of CVD MoS$_2$ capped by ~15 nm AlO$_x$ after heating in air for 10 minutes at (a) 360 ºC, (b) 380 ºC, (c) 400 ºC, and (d) 420 ºC. Red dashed circles mark location of oxidation. At 420 ºC after a few minutes, few μm are oxidized.

## Supporting Information References: